\documentclass[12pt]{article}
\usepackage{latexsym}

\def\ftoday{{\sl  \number\day \space\ifcase\month 
\or Janvier\or F\'evrier\or Mars\or avril\or Mai
\or Juin\or Juillet\or Ao\^ut\or Septembre\or Octobre
\or Novembre \or D\'ecembre\fi
\space  \number\year}}    
\makeatletter
\@addtoreset{equation}{section}
\makeatother

\renewcommand{\a}{\alpha}
\renewcommand{\b}{\beta}
\newcommand{\g}{\gamma}           \newcommand{\G}{\Gamma}
\renewcommand{\d}{\delta}         \newcommand{\D}{\Delta}
\newcommand{\e}{\varepsilon}
\newcommand{\la}{\lambda}        
\newcommand{\m}{\mu}
\newcommand{\n}{\nu}
\newcommand{\om}{\omega}         
              
\renewcommand{\r}{\rho}
\newcommand{\s}{\sigma}           \renewcommand{\S}{\Sigma}
\newcommand{\BB}{{\cal B}}
\newcommand{\CC}{{\cal C}}

\newcommand{\NN}{{\cal N}}

\newcommand{\QQ}{{\cal Q}}
\renewcommand{\SS}{{\cal S}}
\newcommand{\RR}{{\cal R}}

\newcommand{\WW}{{\cal W}}

\newcommand{\sla}{\raise.15ex\hbox{$/$}\kern -.57em} 
\newcommand{\Sla}{\raise.15ex\hbox{$/$}\kern -.70em}

\newcommand{\complex}{{\kern .1em {\raise .47ex
\hbox {$\scriptscriptstyle |$}}
    \kern -.4em {\rm C}}}
\newcommand{\real}{{{\rm I} \kern -.19em {\rm R}}}
\newcommand{\rational}{{\kern .1em {\raise .47ex
\hbox{$\scripscriptstyle |$}}
    \kern -.35em {\rm Q}}}
\renewcommand{\natural}{{\vrule height 1.6ex width
.05em depth 0ex \kern -.35em {\rm N}}}

\newcommand{\half}{\dfrac{1}{2}}

\newcommand{\pad}[2]{{\frac{\partial #1}{\partial #2}}}
\newcommand{\fud}[2]{{\frac{\delta #1}{\delta #2}}}

\newcommand{\dfrac}[2]{{\displaystyle{\frac{#1}{#2}}}}

\newcommand{\ie}{{{\em i.e.},\ }}

\newcommand{\twiddle}{\lower.9ex\rlap{$\kern -.1em\scriptstyle\sim$}}

\newcommand{\equ}[1]{(\ref{#1})}
\newcommand{\eq}{\begin{equation}}
\newcommand{\eqn}[1]{\label{#1}\end{equation}}
\newcommand{\eea}{\end{eqnarray}}
\newcommand{\eqa}{\begin{eqnarray}}
\newcommand{\eqan}[1]{\label{#1}\end{eqnarray}}
\newcommand{\ba}{\begin{array}}
\newcommand{\ea}{\end{array}}
\newcommand{\eqac}{\begin{equation}\begin{array}{rcl}}
\newcommand{\eqacn}[1]{\end{array}\label{#1}\end{equation}}

\begin{document}
%********************************************************
%%\setlength{\baselineskip}{6ex}
%\setlength{\textwidth}{18cm}
%\setlength{\textheight}{26cm}
%%\setlength{\oddsidemargin}{-1cm}
%%\setlength{\evensidemargin}{-1cm}
%\setlength{\topmargin}{-2cm}
%{\large %POUR PREPRINT
%*********************************************************************
\newcommand{\cb}{{\bar c}}
\newcommand{\mn}{{\m\n}}
\newcommand{\pic}{$\spadesuit\spadesuit$}
\newcommand{\?}{{\bf ???}}
\newcommand{\Tr }{\mbox{Tr}\ }
\newcommand{\adot}{{\dot\alpha}}
\newcommand{\bdot}{{\dot\beta}}
\newcommand{\gdot}{{\dot\gamma}}
\titlepage  \noindent
{
%********************************************************
%beta.tex \hspace{1cm} {\bf DRAFT} \hfill{\today}\\

%\hfill{\number\time}\\
%********************************************************
 \noindent
%hep-th/0004048\\
GEF-TH-3/2000 \vspace{8mm}

\noindent
{\bf
{\Huge  Perturbative Beta Function of N=2}} \\

\noindent 
{\bf {\Huge 
Super Yang--Mills  
Theories } }

\vspace{.5cm}
\hrule

\vspace{2cm}

\noindent
{\bf by A.~Blasi$^{* }$, 
V.E.R.~Lemes$^{**}$,
N.~Maggiore$^{*}$,
S.P.~Sorella$^{**}$ ,
A.~Tanzini$^{***}$,
O.S.~Ventura$^{** }$,
L.C.Q.~Vilar$^{**}$}

\noindent
{\footnotesize {\it
$^{*}$ Dipartimento di Fisica -- Universit\`a di Genova --
via Dodecaneso 33 -- I-16146 Genova -- Italy and INFN, Sezione di 
Genova\\
$^{**}$UERJ, Universidade do Estado do Rio de Janeiro,
Departamento de F\'{\i}sica Te\'{o}rica, 
Rua S\~ao Francisco Xavier, 524,
20550-013, Maracan\~{a}, Rio de Janeiro, Brazil\\
$^{***}$Dipartimento di Fisica, Universit\`a di Roma ``Tor 
Vergata'', via della Ricerca Scientifica 1 -- I-00133 Roma -- Italy
} }

\vspace{2cm}
\noindent
{\tt Abstract~:}
An algebraic proof of the nonrenormalization theorem for the 
perturbative beta function of the coupling constant 
of $N=2$ Super Yang--Mills theory is 
provided. The proof relies on a fundamental relationship between the 
$N=2$ Yang--Mills action and the local gauge invariant polynomial 
$\Tr {\phi^{2}}$, $\phi(x)$ being the scalar field of the $N=2$ vector 
gauge multiplet. The nonrenormalization theorem for the $\b_{g}$ 
function follows from the vanishing of the anomalous dimension of 
$\Tr {\phi^{2}}$.

\vfill\noindent
{\footnotesize {\tt PACS codes:}
 03.70.+k (theory of quantized fields), 
 11.10.Gh (renormalization),
 11.10.Hi (renormalization group evolution of parameters),
 11.30.Pb (supersymmetry).
} 
\newpage
%*****************************************************
\section{Introduction}
It has long been known  that  $N=2$ Super Yang--Mills (SYM)
theory displays a set of far-reaching and remarkable 
features, both at the nonperturbative and perturbative level. For 
instance, effects due to instantons can be taken into account in an 
exact way~\cite{Shifman:1999mv,Amati:1988ft}. Needless to say, 
the $N=2$ Yang--Mills 
theory is the corner stone of the duality mechanism discussed by Seiberg 
and Witten~\cite{Seiberg:1994rs} in order to characterize 
the vacuum structure of various supersymmetric gauge models. 

Concerning now the perturbative regime, certainly the most celebrated 
result is the nonrenormalization theorem of the
beta  function of the coupling constant $g$ ($\b_{g}$), 
stating that $\b_{g}$ receives only one-loop contributions. In other 
words, if the $\b_{g}$ function vanishes at the one-loop order, it 
will vanish at all orders of the perturbation theory. 
This unique behavior is usually understood in terms of the analogous 
Adler-Bardeen theorem for the $U(1)$ axial current which, due to 
supersymmetry, belongs to the same supercurrent multiplet of the 
energy-momentum tensor. As a consequence, a deep and strong 
relationship between the $\b_{g}$ function and the coefficient of the 
axial anomaly is expected to hold.
However, up to our knowledge, an algebraic proof of the 
nonrenormalization theorem of the $\b_{g}$ function of $N=2$, based on 
Ward identities and on BRS invariance, is still lacking. 
Indeed the arguments which lead to the exactness of the one--loop 
result for the $\b$-function are based either on a regularization 
procedure which does not respect the supersymmetry or on the higher 
derivative method which has not really been implemented for the $N=2$ 
case. A comprehensive account of the situation is contained in 
Ref.~\cite{West:1990tg}, whose point of view we would like to quote 
({\it ibid.} pag 195)~:
``Here we have stressed these weaknesses not because of a mistrust in 
the arguments for finiteness, but to show that they are not proofs in 
a mathematical sense and that 
there is still room for further work''.
The aim of 
the present work is to fill this gap, providing a purely algebraic 
proof of the aforementioned theorem. 

Before entering into the technical aspects of the paper, it is worth 
spending a few words on the strategy of our proof.
First of all, instead of working with the spinor indices of 
supersymmetry $(\a,\adot)$, we shall make use of the well known 
twisting procedure~\cite{Witten:1988ze}, 
allowing us to replace the 
indices $(\a,\adot)$ with Lorentz vector indices. Of course, the 
physical content of the theory will be left unchanged, for 
the twist is a linear change of variables, and a twisted version of the 
model is perturbatively indistinguishable from the original one. 
However, the use of the twisted variables has the great advantage of 
considerably simplifying the full computation of the relevant BRS 
cohomology classes. In particular, it has been 
possible to prove~\cite{Fucito:1997xm} that the action of the $N=2$ 
Yang--Mills can be related to the local gauge invariant polynomial 
$\Tr {\phi^{2}}$, $\phi(x)$ being the scalar field of the $N=2$ 
multiplet. This important relation will be the essential ingredient 
for the proof of 
the nonrenormalization theorem, as the operator 
$\Tr {\phi^{2}}$ possesses remarkable ultraviolet finiteness 
properties.

Another relevant feature of the twisted formulation is the appearance of 
a scalar supersymmetry charge $\delta$. In fact, the Ward identities 
associated to the scalar supersymmetry transformations play a crucial role
in proving that the anomalous
dimension of $\Tr {\phi^{2}}$ is actually vanishing. Thus, making use
of the relationship between $\Tr {\phi^{2}}$ and the $N=2$ action, we
shall succeed in promoting the ultraviolet finiteness properties of
$\Tr {\phi^{2}}$ to the $\b_{g}$ function of $N=2$, proving therefore
its nonrenormalization theorem.
 
It is worth mentioning that
the twisting procedure has  recently been successfully employed  to
analyze nonperturbative effects, as it allows to understand the $N=2$
Yang--Mills theory in terms of the topological Witten's gauge
theory~\cite{Witten:1988ze}, providing deep insights for the
nonperturbative regime of the model~\cite{Bellisai:2000bc}.
In particular, it has been shown that the remarkable nonrenormalization 
properties
of the holomorphic part of the effective action of the $N=2$ 
SYM~\cite{Seiberg:1988ur}
are ensured in the context of instanton calculus by the Ward identities 
associated to the scalar supersymmetry transformations.
It is interesting to remark that the nonrenormalization properties both of the 
perturbative and the nonperturbative sector of the $N=2$ SYM theory
can be related to the Ward identities associated to the scalar supersymmetry
present in its twisted formulation.

The organization of the paper is as follows. Section~2 is devoted to a 
brief review of the twisting procedure for $N=2$ supersymmetry. In 
section~3 the quantization of the theory in the Wess-Zumino gauge is 
performed following the BRS 
procedure~\cite{Fucito:1997xm,White:1992ai,Maggiore:1995xw}. In 
section~4 the relationship between the gauge invariant polynomial $\Tr 
{\phi^{2}}$ and the $N=2$ Yang--Mills action will be presented. 
Section~5 will deal with the ultraviolet finiteness properties of 
$\Tr {\phi^{2}}$. Finally, in section~6 the algebraic proof of the 
nonrenormalization theorem for the $\b_{g}$ function of $N=2$ 
Yang--Mills will be given. 

%*****************************************************

%*****************************************************
\section{The Twist }\label{sez2}
The starting key of addressing the problem is to work with the 
twisted version of $N=2$  SYM theories, which, in 
the Wess--Zumino (WZ) gauge, coincide with Topological Yang--Mills 
(TYM) 
theory. In this section, we sketch the basics of the twisting 
procedure, originally introduced by Witten in~\cite{Witten:1988ze} 
and fully 
developed for $N=2$ SYM in~\cite{Fucito:1997xm,Marino:1996sd}.

The $N=2$ susy algebra reads
\eqa
\{\QQ^{i}_{\a},\overline{\QQ}_{j\adot}\}&=&                                      
\d^{i}_{j}(\s^{\m})_{\a\adot}\partial_{\m} 
\nonumber \\
\{\QQ^{i}_{\a},\QQ^{j}_{\b}\} &=& 
                   \{\overline{\QQ}_{i\adot},\overline{\QQ}_{j\bdot}\} = 0\ ,
\eqan{n=2algebra}
where $(\QQ^{i}_{\a},\overline{\QQ}_{j\adot})$ are the supersymmetry 
charges, 
indexed by $i=1,2$ and Weyl spinor indices $\a,\adot = 1,2$. The 
total 
number of supercharges is therefore eight. The simple observation 
that the 
indices $(i,\a)$ both run from one to two, suggests the idea of 
identifying the supersymmetry index with the spinor index
\eq
i \equiv \a \ ,
\eqn{1}
operation which amounts to a modification of the rotation group of 
the 
theory
\eq
SU(2)_{L} \times SU(2)_{R} \longrightarrow SU(2)_{L} \times 
SU(2)_{R}^{\prime}\ ,
\eqn{2}
where $SU(2)_{R}^{\prime}$ is the diagonal sum of the original  
$SU(2)_{R}$ and  $SU(2)_{I}$, the group of transformations of the 
supersymmetry index $i$. The global symmetry group of $N=2$ SYM 
finally is
\eq
SU(2)_{L} \times SU(2)_{R}^{\prime}\times U(1)_{R}\ ,
\eqn{3}
where $U(1)_{R}$ is associated to the ${\cal R}$--symmetry, according 
to which the charges $\QQ^{i}_{\a}$ and $\overline{\QQ}_{i\adot}$ are 
assigned eigenvalues $+1$ and $-1$ respectively.

The twisted supercharges are 
\eqa
\QQ^{i}_{\a} &\longrightarrow& \QQ^{\b}_{\a} \nonumber \\
\overline{\QQ}_{i\adot} &\longrightarrow& \overline{\QQ}_{\a\adot}\ .
\eqan{4}
The procedure can then be pushed further to the point of getting rid 
of the spinor index. Indeed, $\QQ^{\b}_{\a}$ and 
$\overline{\QQ}_{\a\adot}$ 
can be rearranged as follows
\eqa
\delta &\equiv& \frac{1}{\sqrt{2}} \varepsilon^{\a\b}\QQ_{\a\b} 
\label{delta}\\
\delta_{\m} &\equiv& \frac{1}{\sqrt{2}} 
                       \overline\QQ_{\a\adot}(\s^{\m})^{\adot\a}  
                       \label{deltamu}\\
\delta_{\m\n} &\equiv& \frac{1}{\sqrt{2}} 
(\s_{\m\n})^{\a\b}\QQ_{\b\a}\ ,\label{deltamunu} 
\eqan{5}
where we adopted the usual conventions for the quantities 
$\varepsilon^{\a\b}$, $(\s^{\m})^{\adot\a}$ and 
$(\s_{\m\n})^{\a\b}$~\cite{Bagger:1990qh}. 
Notice that, due to the properties of $\s_{\m\n}$, the operator 
$\d_{\m\n}$ is selfdual~:
\eq
\d_{\m\n} = \frac{1}{2} \varepsilon_{\m\n\r\s} \d^{\r\s}\ ,
\eqn{6}
reducing to three the number of its independent components.

The familiar supersymmetry algebra~\equ{n=2algebra} correspondingly 
twists into
\eqa
\d^{2} &=& 0 \nonumber \\
\{\d,\d_{\m}\} &=& \partial_{\m} \label{subalgebra}\\
\{\d_{\m},\d_{\n}\} &=& 0 \nonumber
\eqan{7}
\eqa
\{\d_{\m},\d_{\r\s}\} &=& - (\varepsilon_{\m\n\r\s}\partial^{\n} + 
g_{\m\r}\partial_{\s} - g_{\m\s}\partial_{\r}) \nonumber \\
\{\d_{\m\n},\d\} &=& \{\d_{\m\n},\d_{\r\s}\} =0
\eqan{else}
A key feature of the twisted algebra is the appearance of the
scalar supersymmetry charge $\delta$, which is still an invariance of the
theory when this is formulated on a generic (differential) manifold
$M$. We recall in fact that in Witten's Topological Yang--Mills theory the 
observables are {\it defined} as cohomology classes of $\delta$,
which is thus treated in this context as a BRS--like operator 
\footnote{Actually since the 
Witten's theory is formulated in the Wess--Zumino gauge, 
one is led to analyze the equivariant cohomology of the
corresponding BRS operator, as observed in~\cite{stora}}.
The corresponding correlation functions are non vanishing in the nonperturbative
sector of the theory and can can be indeed evaluated by a semiclassical 
expansion around  an instanton background. They are independent of the metric
on $M$ by virtue of the Ward identities associated to $\delta$,
and are related with the Donaldson invariants \cite{Witten:1988ze}. 
The underlying topological nature of the twisted theory is already 
revealed by the subalgebra~\equ{subalgebra}, which allows to write the 
space--time derivative as a $\delta$--exact term. Notice that this 
subalgebra is a common feature 
of all known topological quantum field 
theories~\cite{Delduc:1989ft,Piguet:1995er}.

The price of working in the Wess--Zumino gauge, is the appearance in 
the r.h.s. of~\equ{n=2algebra} (and hence of~\equ{subalgebra} 
and~\equ{else}) of breakings, 
which take the form of field dependent gauge transformations and 
equations of motion. The drawbacks deriving from those breakings for 
the quantization of the theory have been stressed by Breitenlohner 
and 
Maison in~\cite{Breitenlohner:1985kh}.

The fields of $N=2$ WZ gauge multiplet 
$(A_{\m},\psi^{i}_{\a},\overline{\psi}^{i}_{\adot},\phi,\overline\phi)$, 

belonging to the adjoint representation of the gauge group, after the 
twisting procedure become
\eq
(A_{\m},\psi_{\m},\chi_{\m\n},\eta,\phi,\overline\phi)\ .
\eqn{8}
The twist, acting on the internal index $i$, leaves the gauge 
connection $A_{\m}$ and the scalars $(\phi,\overline\phi)$ unaltered, 
while the spinors $(\psi^{i}_{\a},\overline{\psi}^{i}_{\adot})$ turn 
into
\eqa
\psi^{i}_{\b} &\rightarrow & = \frac{1}{2} ( \psi_{(\a\b)} + 
                                             \psi_{[\a\b]}) \\ 
\overline\psi^{i}_{\adot} &\rightarrow & \overline\psi_{\a\adot} 
\rightarrow \psi_{\m} = 
(\overline\s_{\m})^{\a\adot}\overline\psi_{\a\adot}					     
\eqan{9}
and
\eqa
\psi_{(\a\b)} &\rightarrow& \chi_{\m\n} = 
(\s_{\m\n})^{\a\b}\psi_{(\a\b)} \\
\psi_{[\a\b]} &\rightarrow& \eta = \varepsilon^{\a\b}\psi_{[\a\b]}\ .
\eqan{10}
Notice that $(\psi_{\m},\chi_{\m\n},\eta)$ anticommute, due to their 
spinor origin.

\newpage

Finally, twisting the action of $N=2$ SYM, remarkably yields the TYM 
action~\cite{Witten:1988ze}
\eq
\SS^{N=2}_{SYM}
(A_{\m},\psi^{i}_{\a},\overline{\psi}^{i}_{\adot},\phi,\overline\phi)
\longrightarrow 
\SS_{TYM}
(A_{\m},\psi_{\m},\chi_{\m\n},\eta,\phi,\overline\phi)
\eqn{11}
and
\begin{eqnarray}
\mathcal{S}_{TYM} &=&\frac 1{g^2}\Tr\displaystyle\int d^4x\;\left( 
\frac
12F_{\mu \nu }^{+}F^{+\mu \nu }\;-\chi ^{\mu \nu }(D_\mu \psi _\nu 
-D_\nu
\psi _\mu )^{+}\;\right.  \label{tym} \\
&&\;\;\;\;\;\;\;\;\;\;\;\;\;\;\;\;\;+\eta D_\mu \psi ^\mu \;-\frac 12%
\overline{\phi }D_\mu D^\mu \phi \;+\frac 12\overline{\phi }\left\{ 
\psi
^\mu ,\psi _\mu \right\}  \nonumber \\
&&\;\;\;\;\;\;\;\;\;\;\;\;\left. \;\;\;-\frac 12\phi \left\{ \chi 
^{\mu \nu
},\chi _{\mu \nu }\right\} \;-\frac 18\left[ \phi ,\eta \right] \eta 
-\frac
1{32}\left[ \phi ,\overline{\phi }\right] \left[ \phi ,\overline{\phi 
}%
\right] \right) \;,  \nonumber
\label{stym}\end{eqnarray}
where
\begin{eqnarray}
&&F_{\mu \nu }^{+} = F_{\mu \nu }+\frac 12\varepsilon _{\mu \nu \rho 
\sigma
}F^{\rho \sigma }\;,\;\;\;\;\;\;\widetilde{F}_{\mu \nu }^{+}=\frac
12\varepsilon _{\mu \nu \rho \sigma }F^{+\rho \sigma }=F_{\mu \nu 
}^{+}\;,
\label{f+} \\
&&(D_\mu \psi _\nu -D_\nu \psi _\mu )^{+}=(D_\mu \psi _\nu -D_\nu 
\psi _\mu
)+\frac 12\varepsilon _{\mu \nu \rho \sigma }(D^\rho \psi ^\sigma 
-D^\sigma
\psi ^\rho )\;.  \label{Dpsi}
\end{eqnarray}
Notice that the theory has the unique coupling constant $g^{2}$, 
which 
in this parametrization is easily related to the number $L$ of loops 
in the Feynman diagrams computations~:
\eq
L\ \mbox{loops} \longleftrightarrow (g^{2})^{L-1}\ .
\eqn{12}
The action $\mathcal{S}_{TYM}$ \equ{tym} is left invariant by the 
twisted supersymmetry charges
\eq
\d\ \mathcal{S}_{TYM} = \d_{\m}\ \mathcal{S}_{TYM}=\d_{\m\n}\ 
\mathcal{S}_{TYM} =0\ ,
\eqn{13}
and it is also gauge invariant, \ie
\eq
\d_{gauge}\ \mathcal{S}_{TYM} =0 \ .
\eqn{14}
We stress that the twist simply corresponds, on the flat 
Euclidean spacetime ${\real}^4$,
to a linear change of variables, 
and 
therefore the twisted theory is perturbatively indistinguishable from 
the original one. In particular, all the perturbative results are 
the same for the two theories. This fact has been already verified 
under many circumstances: we just recall here the $algebraic$ absence 
of gauge anomalies~\cite{noanom}, 
which is quite uncommon 
for four dimensional gauge 
field theories. Indeed, in these cases it usually happens that the 
gauge 
anomaly is algebraically allowed, and only the vanishing of its 
coefficient may be invoked to recover perturbative renormalization.
Another quantity which has been verified to go over unchanged to the 
twisted related theories is the value of the 1--loop $\b$ function of 
the coupling $g$, which is universal, \ie 
scheme--independent~\cite{1loopbeta}. What is still missing, is the 
main aim of the present paper, namely the algebraic, regularization 
independent proof of the celebrated vanishing above 1 loop of the 
$\b_{g}$ function of $N=2$ SYM theories, and hence of TYM theory. 
This means, according to the parametrization adopted for  the action 
$\mathcal{S}_{TYM}$ \equ{tym}, that  $\b_{g}$ is proportional to 
$g^{3}$, to all orders of perturbation theory
\eq
\b_{g} \propto g^{3}\ .
\eqn{15}
We will be able in the next sections to precise our claim \equ{15}, 
which 
makes sense only after a particular renormalization scheme has been 
adopted, since it is evident that a change of scheme generally 
induces in the 
1 loop only $\b$ function terms of arbitrarily high power of the 
coupling constant. We address the interested reader to the evergreen 
lectures given by D.J.Gross~\cite{Gross:1975vu}. 
%*****************************************************

%*****************************************************
\section{Quantum Extension}\label{sez3}

For what follows, it is useful to briefly illustrate the technique 
employed for the quantum extension of the model. As already said, the 
way in which the algebraic structure is broken in the Wess--Zumino 
gauge, is incompatible with the construction of a quantum vertex $\G$ 
satisfying quantum symmetries (gauge and 
supersymmetry)~\cite{Breitenlohner:1985kh}. The 
puzzle has been solved 
in~\cite{Maggiore:1995dw,White:1992ai,Maggiore:1995xw}, where the 
renormalization of $N=2$ SYM has been performed, and the same 
procedure has been successfully applied in~\cite{Fucito:1997xm} for 
TYM. 
The idea is to collect all the relevant symmetries of the theory into 
an unique operator, by means of some global ghosts. Here, by relevant 
symmetries we mean the ordinary BRS symmetry $s$, and the symmetries 
of the topological subalgebra~\equ{subalgebra}, the charges 
$\d_{\m\n}$ being redundant in order to define the 
theory~\cite{Fucito:1997xm}. Let us therefore define the extended BRS 
operator
\begin{equation}
\mathcal{Q=}\ s+\omega \delta +\varepsilon ^\mu \delta _\mu
\;,  \label{Q-op}
\end{equation}
where $\om$ and $\varepsilon^{\m}$ are global ghosts. The action of 
$\QQ$ on the fields belonging to the twisted $N=2$ gauge 
supermultiplet is
\begin{eqnarray}
\mathcal{Q}A_\mu &=&-D_\mu c+\omega \psi _\mu +\frac{\varepsilon ^\nu 
}2\chi
_{\nu \mu }+\frac{\varepsilon _\mu }8\eta 
\label{Q-transf} \\
\mathcal{Q}\psi _\mu &=&\left\{ c,\psi _\mu \right\} -\omega D_\mu 
\phi
+\varepsilon ^\nu \left( F_{\nu \mu }-\frac 12F_{\nu \mu }^{+}\right) 
-\frac{%
\varepsilon _\mu }{16}[\phi ,\overline{\phi }]  \nonumber \\
\mathcal{Q}\chi _{\sigma \tau } &=&\left\{ c,\chi _{\sigma \tau 
}\right\}
+\omega F_{\sigma \tau }^{+}+\frac{\varepsilon ^\mu }8(\varepsilon 
_{\mu
\sigma \tau \nu }+g_{\mu \sigma }g_{\nu \tau }-g_{\mu \tau }g_{\nu 
\sigma
})D^\nu \overline{\phi }\;  \nonumber \\
\mathcal{Q}\eta &=&\left\{ c,\eta \right\} +\frac \omega 2[\phi 
,\overline{%
\phi }]+\frac{\varepsilon ^\mu }2D_\mu \overline{\phi }\nonumber \\
\mathcal{Q}\phi &=&\left[ c,\phi \right] -\varepsilon ^\mu \psi _\mu 
\nonumber \\
\mathcal{Q}\overline{\phi } &=&\left[ c,\overline{\phi }\right] 
+2\omega
\eta  \ .\nonumber \\
\end{eqnarray}
Moreover, $\mathcal{Q}$
is defined on the $\Phi\Pi$ -- multiplet ghost--antighost--Lagrange 
multiplier $(c,\bar{c},b)$ as follows~:
\begin{eqnarray}
\mathcal{Q}c &=&c^2-\omega ^2\phi -\omega \varepsilon ^\mu A_\mu 
+\frac{%
\varepsilon ^2}{16}\overline{\phi }  \nonumber\\
\mathcal{Q}\overline{c}\; &=&b
 \\
\mathcal{Q}b &=&\omega \varepsilon ^\mu \partial _\mu \overline{c}\ 
.  \nonumber
\label{Q-ext}\end{eqnarray}
Besides describing a symmetry of the action $\SS_{TYM}$, 
the relevant property of the operator $\mathcal{Q}$ is~:
\eq
\QQ^{2} = \om\varepsilon^{\m}\partial_{\m} + \ \mbox{eqs. of motion}\ 
,
\eqn{nilpot}
which allows for a standard application of the BRS technique, since 
the 
extended BRS operator $\QQ$ is on--shell nilpotent in the space of 
integrated local functionals.

The complete classical action is~\cite{Fucito:1997xm}
\eq
\S = \SS_{TYM} + S_{gf} + S_{ext} + S_{\Box}\ ,
\eqn{class-ax}
where
\eq
S_{gf} =\QQ\int d^4x\;\Tr(\overline{c}\partial A)  
\eqn{landau-g-fi}
is the (Landau) gauge fixing term;
\begin{eqnarray}
\mathcal{S}_{ext} &=&\displaystyle \Tr\int 
d^4x\;(\;L\mathcal{Q}c+D\mathcal{Q}%
\phi +\Omega ^\mu \mathcal{Q}A_\mu +\xi ^\mu \mathcal{Q}\psi _\mu \;
\label{s-ext} \\
&&\;\;\;\;\;\;\;\;\;\;\;\;\;\;\;\;\;\;\;+\rho 
\mathcal{Q}\overline{\phi }%
+\tau \mathcal{Q}\eta +\frac 12B^{\mu \nu }\mathcal{Q\chi }_{\mu \nu 
}\;)\;,
\nonumber
\end{eqnarray}
couples external sources (called  antifields in the 
Batalin--Vilkovisky formalism) \\
$(L,D,\Omega ^\mu
,\xi ^\mu ,\rho ,\tau ,B_{\mu \nu }),$ to the nonlinear variations of 
the quantum fields 
\\
$(c,\phi,A_{\m},\psi_{\m},\overline\phi,\eta,\chi_{\m\n})$ 
respectively; finally
\begin{equation}
\mathcal{S}_{\Box}=\frac{g^{2}}{4}\Tr \int d^4x\left( \frac{1}{2}
\omega ^2B^{\mu \nu}
B_{\mu \nu }-\omega B^{\mu \nu }\varepsilon _\mu \xi _\nu
-\frac{1}{8}\varepsilon ^\mu \varepsilon ^\nu \xi _\mu \xi _\nu 
+\frac{1}{8}\varepsilon ^2\xi ^2\right) \;,  \label{s-quad}
\end{equation}
is introduced in order to take care of the fact that $\QQ$ is 
nilpotent (on integrated functionals) once the equations of motion 
are 
used.

The action $\S$ \equ{class-ax} satisfies 
an extended Slavnov--Taylor (ST) identity
\begin{equation}
\mathcal{S}(\Sigma )\;\mathcal{=\;}\om\e^{\m}\D^{cl}_{\m}\;,  
\label{tym-s-t}
\end{equation}
where
\begin{eqnarray}
\mathcal{S}(\Sigma ) &=&\Tr\int d^4x\left( \frac{\delta \Sigma 
}{\delta A^\mu 
}\frac{\delta \Sigma }{\delta \Omega _\mu }+\frac{\delta \Sigma 
}{\delta \xi
^\mu }\frac{\delta \Sigma }{\delta \psi _\mu }+\frac{\delta \Sigma 
}{\delta L%
}\frac{\delta \Sigma }{\delta c}+\frac{\delta \Sigma }{\delta 
D}\frac{\delta
\Sigma }{\delta \phi }\;+\frac{\delta \Sigma }{\delta \rho 
}\frac{\delta
\Sigma }{\delta \overline{\phi }}\right.    \nonumber  \\
&&\;\;\;\;\;\;\;\;\;\;\;\;\;\;+\frac{\delta \Sigma }{\delta \tau 
}\frac{%
\delta \Sigma }{\delta \eta }+\frac 12\frac{\delta \Sigma }{\delta 
B^{\mu
\nu }}\frac{\delta \Sigma }{\delta \chi _{\mu \nu }}\;\;
+b \frac{\delta \Sigma }{\delta \overline{c}}  %\nonumber \\
%&&\;\;\;\;\;\;\;\;\;\;\;\;\;\;
\left. 
+\omega \varepsilon ^\mu \partial _\mu 
\overline{c}\frac{\delta \Sigma }{\delta b}\right)
\;.\label{tym-s-t-op}
\end{eqnarray}
In the above expression \equ{tym-s-t} for the extended ST identity, 
$\D^{cl}_{\m}$ is an integrated linear polynomial in the quantum 
fields~:
\eq
\Delta _\mu ^{cl} =\Tr\int d^4x(\;L\partial _\mu c-D\partial _\mu \phi
-\Omega ^\nu \partial _\mu A_\nu +\xi ^\nu \partial _\mu \psi _\nu \; 
-\rho \partial _\mu \overline{\phi }+\tau
\partial _\mu \eta +\frac 12B^{\nu \sigma }\partial _\mu \chi _{\nu 
\sigma
}\;)\; ;  \label{v-break}
\end{equation}
therefore,  $\D^{cl}_{\m}$ does not get quantum corrections and the 
breaking of the ST identity \equ{tym-s-t} is only $classical$.

The whole algebra is summarized into
\begin{equation}
\mathcal{B}_{\Sigma }\mathcal{B}_{{\Sigma }}=\omega
\varepsilon ^\mu \mathcal{P}_\mu \;,  \label{nil-lin-tym}
\end{equation}
where
\begin{eqnarray}
\mathcal{B}_{{\Sigma }} &=&\Tr\int d^4x\left( \frac{\delta {%
\Sigma }}{\delta A^\mu }\frac \delta {\delta \Omega _\mu 
}+\frac{\delta 
{\Sigma }}{\delta \Omega _\mu }\frac \delta {\delta A^\mu }+\frac{%
\delta {\Sigma }}{\delta \psi _\mu }\frac \delta {\delta \xi ^\mu }+%
\frac{\delta {\Sigma }}{\delta \xi ^\mu }\frac \delta {\delta \psi
_\mu }+\frac{\delta {\Sigma }}{\delta L}\frac \delta {\delta
c}\right)  \nonumber   \\
&&\;\;\;\;\;\;\;\;\;+\frac{\delta {\Sigma }}{\delta c}\frac \delta
{\delta L}+\frac{\delta {\Sigma }}{\delta \phi }\frac \delta {\delta
D}+\frac{\delta {\Sigma }}{\delta D}\frac \delta {\delta \phi }+%
\frac{\delta {\Sigma }}{\delta \overline{\phi }}\frac \delta {\delta
\rho }+\frac{\delta {\Sigma }}{\delta \rho }\frac \delta {\delta 
\overline{\phi }}+\frac{\delta {\Sigma }}{\delta \eta }\frac \delta
{\delta \tau }  \label{n-tym-lin-op} \\
&&\;\;\;\;\;\;\;\;\;\left. +\frac{\delta {\Sigma }}{\delta \tau }%
\frac \delta {\delta \eta }+\frac 12\frac{\delta {\Sigma }}{\delta
\chi _{\mu \nu }}\frac \delta {\delta B^{\mu \nu }}+\frac 
12\frac{\delta 
{\Sigma }}{\delta B^{\mu \nu }}\frac \delta {\delta \chi _{\mu \nu
}}+b\frac \delta {\delta \overline{c}}+\omega \varepsilon ^\mu 
\partial _\mu 
\overline{c}\frac \delta {\delta b}\right) \;,  \nonumber
\end{eqnarray}
is the linearized extended ST operator, and
\begin{eqnarray}
\mathcal{P}_\mu  &=&{\sum }_{i}\int d^4x\left( \partial _\mu
\varphi ^i\frac{\delta }{\delta \varphi ^i}+\partial _\mu \varphi
^{*i}\frac{\delta  }{\delta \varphi ^{*i}}\right) =\;0\;,\;\;
\label{transl-inv} \\
\;\;\varphi ^i &=&\mathrm{all\;the\;fields\;(}A,\psi ,\phi 
,\overline{\phi }%
,\eta ,\chi ,c,\overline{c},b\mathrm{)\;},  \nonumber \\
\varphi ^{*i} &=&\mathrm{all\;the\;antifields\;(}\Omega ,\xi 
,L,D,\rho ,\tau
,B\mathrm{)}\;.\nonumber
\end{eqnarray}
is the functional generator of  spacetime translations.

In \cite{Fucito:1997xm}, the renormalization program has been 
completed 
by showing that the action $\S$ is stable under radiative corrections 
and that no anomalies occur, so that the classical relation 
\equ{tym-s-t} can safely be implemented for the 1PI generating 
functional 
\eq
\G = \S + O(\hbar)\ ,
\eqn{gamma}
\ie
\eq
\SS(\G) = \om\e^{\m}\D^{cl}_{\m}\ .
\eqn{quantumst}

Let us close this section by summarizing  the quantum numbers of the 
fields and parameters entering the theory.
\begin{equation}
\stackrel{Quantum\;numbers: \ Quantum\ Fields}{
\begin{tabular}{|c|c|c|c|c|c|c|}
\hline
& $A_\mu $ & $\chi _{\mu \nu }$ & $\psi _\mu $ & $\eta $ & $\phi $ & 
$%
\overline{\phi }$ \\ \hline
$\dim \mathrm{.}$ & $1$ & $3/2$ & $3/2$ & $3/2$ & $1$ & $1$ \\ \hline
$\mathcal{R}-\mathrm{ch}\arg \mathrm{e}$ & $0$ & $-1$ & $1$ & $-1$ & 
$2$ & $%
-2$ \\ \hline
$\mathrm{gh-number}$ & $0$ & $0$ & $0$ & $0$ & $0$ & $0$ \\ \hline
$\mathrm{nature}$ & $comm.$ & $ant.$ & $ant.$ & $ant.$ & $comm.$ & 
$comm.$
\\ \hline
\end{tabular}
}  \label{fields-table}
\end{equation}

\begin{equation}
\stackrel{Quantum\;numbers:\ Ghosts}{
\begin{tabular}{|c|c|c|c|c|}
\hline
& $c$&$\bar{c}$ & $\omega $ & $\varepsilon _\mu $ \\ \hline
$\dim .$ & $0$&$2$ & $-1/2$ & $-1/2$ \\ \hline
$\mathcal{R}-\mathrm{ch}\arg \mathrm{e}$ & $0$&$0$ & $-1$ & $1$ \\ 
\hline
$\mathrm{gh-number}$ & $1$&$-1$ & $1$ & $1$ \\ \hline
$\mathrm{nature}$ & $ant.$&$ant.$ & $comm.$ & $comm.$ \\ \hline
\end{tabular}
}  \label{ghosts-table}
\end{equation}

\begin{equation}
\stackrel{Quantum\;numbers:\ External \ Sources}{
\begin{tabular}{|c|c|c|c|c|c|c|c|}
\hline
& $L$ & $D$ & $\gamma ^\mu $ & $\xi ^\mu $ & $\rho $ & $\tau $ & 
$B^{\mu \nu
}$ \\ \hline
$\dim .$ & $4$ & $3$ & $3$ & $5/2$ & $3$ & $5/2$ & $5/2$ \\ \hline
$\mathcal{R}-\mathrm{ch}\arg \mathrm{e}$ & $0$ & $-2$ & $0$ & $-1$ & 
$2$ & $%
1 $ & $1$ \\ \hline
$\mathrm{gh-number}$ & $-2$ & $-1$ & $-1$ & $-1$ & $-1$ & $-1$ & $-1$ 
\\ 
\hline
$\mathrm{nature}$ & $comm.$ & $ant.$ & $ant.$ & $comm.$ & $ant.$ & 
$comm.$ & 
$comm.$ \\ \hline
\end{tabular}
}  \label{antifields-table}
\end{equation}
%*****************************************************

%*****************************************************
\section{The Action, and its Relation with  $\Tr\phi^{2}$}\label{sez4}

By expanding the linearized ST operator \equ{n-tym-lin-op} 
in powers of the global ghost $\e^{\m}$, we have
\begin{equation}
\mathcal{B}_{\S}=b_{\S}%
+\varepsilon ^\mu \mathcal{W}_\mu +\frac 12\varepsilon ^\mu 
\varepsilon
^\nu \mathcal{W}_{\mu \nu }\;,  \label{lin-dec}
\end{equation}
and the algebraic relation \equ{nil-lin-tym} implies that the 
operators $b_{\S}$ and $\mathcal{W}_\mu$ must obey
\eqa
b_{\S}b_{\S}&=&0 
\label{b-exact-nilp} \\
\left\{ b_{\S},\mathcal{W}_\mu \right\} &=&\omega 
\mathcal{P}_\mu \;.  
\eqan{b-W-dec}
As it is well known \cite{Piguet:1995er,Dixon:1991wi}, 
a theorem assures that the 
integrated cohomology of $\BB_{\S}$ is isomorphic to a subspace of 
the 
integrated cohomology of the first term in the expansion 
\equ{lin-dec}, \ie $b_{\S}$, whose local cohomology is known 
\cite{Delduc:1996yh} 
to consist of invariant polynomials in the scalar field $\phi$
\begin{equation}
\mathcal{P}_n(\phi )=\Tr\left( \frac{\phi ^n}n\right) 
\;,\;\;\;\;\;\;n\geq
2\;.  \label{b-cohomology}
\end{equation}

Now, the polynomials $\mathcal{P}_n(\phi )$ become cohomologically 
trivial if we relax the condition of analyticity in the global ghost 
$\om$. In fact, for instance, 
\begin{equation}
\mathcal{P}_2(\phi )\;=\frac 12\Tr\phi ^2 \;=\frac 12
b_{\S}\; \Tr\left( -\frac 1{\omega ^2}c\phi
+\frac 1{3\omega ^4}c^3\right) \;.
 \label{p2-inv-pol}
\end{equation}
This means that the local cohomology of the operator $b_{\S}$, and 
hence the integrated cohomology of the full operator $\BB_{\S}$, is 
trivial if we take into account also field polynomials which {\it are 
not} 
analytic in the parameter $\om$. On the other way, we have to 
remember 
that the physical content of TYM theory is not empty, being given by 
the non empty cohomology of $\BB_{\S}$, with the requirement of 
analyticity in $\om$, or by the equivariant cohomology of the 
ordinary BRS operator, as defined in \cite{stora}, where TYM 
is seen as a topological quantum field theory of the Witten type, \ie 
generated entirely by a BRS cocycle. More on this can be found in 
\cite{Fucito:1997xm}. Thus, the requirement of analyticity in $\om$ 
is 
not just a matter of convenience, but it is founded in the quantum 
field theory rules. However, in perturbation theory the diagrammatic 
expansion of the generating functional $\G$ is obviously analytic in 
the parameters of the theory. In other words, counterterms for the 
action must be analytic in $\om$, while this is not necessarily so 
for other elements of the quantum theory, like for instance quantum 
insertions of the type $\pad{\G}{\a}$, where $\a$ is any parameter of 
the theory. Indeed, the possible poles can be easily removed from 
$\pad{\G}{\a}$ by a multiplication for a suitable power of $\om$.
 
Turning back to the original issue, in the space of integrated local 
functionals of vanishing $\RR$-charge, zero ghost number and 
canonical dimension four, the cohomology of $\BB_{\S}$ is given by 
the 
unique element~\cite{Fucito:1997xm}
\begin{equation}
\varepsilon ^{\mu \nu \rho \tau }\mathcal{W}_\mu \mathcal{W}_\nu 
\mathcal{W}%
_\rho \mathcal{W}_\tau \int d^4x\;\Tr\left( \frac{\phi 
^{2\;}}2\right) \;,
\label{nontr-el}
\end{equation}
where $\mathcal{W}_\mu$ is the operator defined by the filtration 
of $\BB_{\S}$ in \equ{lin-dec}. It is clear from \equ{nontr-el} that 
the role of the operator $\WW_{\m}$ is that of extracting the 
integrated cohomology classes from the non-integrated ones, in 
perfect 
analogy with what happens in topological field theories, which are 
all characterized by the algebraic relation \equ{b-W-dec}, according 
to 
which the spacetime derivative is written as an anticommutator 
between a BRS operator and a vectorial 
supersymmetry~\cite{Delduc:1989ft,Piguet:1995er}. 
The operator $\WW_{\m}$ acts therefore as a climbing 
up operator for the descent 
equations, and the non trivial part of the action is obtained by 
repeated applications of $\WW_{\m}$ to the lowest term of the ladder 
\cite{Piguet:1995er}. 
For our concern, the twisted action \equ{class-ax} of $N=2$ 
SYM can be rewritten, modulo a trivial $\BB_\S$--cocycle, as 
\eq
\S \approx -\frac{1}{3g^{2}} \WW^{4} \int d^4x\;\Tr 
\frac{\phi^{2}}{2} 
%+ \BB_{\S}\; \int d^{4}x\; \Tr \left( \phi D - \xi^{\m}\psi_{\m}\right )
\ ,
\eqn{w4action}
where
\eq
\WW^{4} \equiv \e^{\m\n\r\s}\WW_{\m}\WW_{\n}\WW_{\r}\WW_{\s}\ .
\eqn{w4}

The relation \equ{w4action} is of fundamental importance for what 
follows. Although it is self-explanatory, it is worth to stress that 
the bulk of the theory is the composite operator 
$\Tr \frac{\phi^{2}}{2}$, which somehow contains all the information 
on the 
physics of the theory. The operator $\WW_{\m}$ is nothing more than a 
BRS decomposition of the spacetime derivative. In this respect, 
$\Tr \frac{\phi^{2}}{2}$ can be regarded as a sort of prepotential of 
$N=2$ SYM (and of TYM, too), at least in this twisted version. Notice 
also that $\phi(x)$ is just the scalar field of the original, 
untwisted, theory, namely the scalar of the gauge multiplet of $N=2$ 
SYM, and that the composite operator $\Tr \frac{\phi^{2}}{2}$ is 
already known to play a decisive role for the exact solution of the 
theory. In fact, it is precisely its v.e.v. $<\Tr 
\frac{\phi^{2}}{2}> = u$ which parametrizes the gauge non equivalent 
vacua of the theory \cite{Seiberg:1994rs}. 
Here, it is instructive to recover the 
key role of this composite operator in a rather different context. 
Intuitively, it is also clear what will  our strategy be for the 
following: the nonrenormalization properties of the full action are 
closely related to those of its ``generating kernel'' 
$\Tr \frac{\phi^{2}}{2}$, and on this issue we will concentrate in 
the 
next sections.
%*******************************************************

%*****************************************************
\section{Non Renormalization of Tr $\phi^{2}$}\label{sez5}

In order to show that the composite operator $\Tr \frac{\phi^{2}}{2}$
has vanishing anomalous dimensions, the ordinary BRS and 
$\d$--symmetry \equ{delta} are sufficient.
Therefore, from now on throughout the rest of this paper, we shall 
put 
$\varepsilon^\m\equiv 0$, and we shall write
$\S \equiv \left. \S\right|_{\varepsilon =0}$,
$\G \equiv \left. \G\right|_{\varepsilon =0}$ and
$\SS(\G) \equiv \left. \SS(\G)\right|_{\varepsilon =0}$,
while the linearized ST operator will be that 
defined in~\equ{lin-dec} by
$\left.\BB_\S\right|_{\varepsilon =0} = b_\S$\ .

Let us consider
\eq
\QQ^{\prime} \equiv  s +\om\d\ ,
\eqn{qprime}
which, like $\QQ$ in \equ{Q-op}, describes a symmetry of $\SS_{TYM}$, 
and it is nilpotent on shell.

Consider the composite operators
\begin{eqnarray}
    \Omega_{\phi^{2}} &\equiv& \Tr \om^{4}\phi^{2} \label{omfiquad} \\
    \Omega_{c^{3}} &\equiv& \Tr 
    (-\om^{2}c\phi+\frac{1}{3}c^{3})\label{omcicub}\ .
\end{eqnarray}
The following relation holds
\eq
\Omega_{\phi^{2}} = \QQ^{\prime} \; \Omega_{c^{3}}\ .
\eqn{rel}
The action
\eq
\S^{\prime} = \S + S_{\s\la}\ ,
\eqn{axprime}
where 
\eq
S_{\s\la} = \int d^{4}x\; (\s \Omega_{\phi^{2}} + \la 
\Omega_{c^{3}})\ ,
\eqn{slm}
$\left(\s(x),\la(x)\right)$ being external sources, 
satisfies the ST identity
\eq
\SS^{\prime}(\S^{\prime}) = 0\ ,
\eqn{stidprime}
where
\eq
\SS^{\prime}(\S^{\prime}) = \SS(\S^{\prime}) + \int d^{4}x\; 
\la\fud{\S^{\prime}}{\s}\ .
\eqn{stprime}
It is apparent from the expression \equ{stprime} that the external 
fields $\s(x)$ and $\la(x)$, transforming one into the other, form a 
BRS doublet. This means that the cohomology of the linearized 
ST operator $b_{\S^{\prime}}$ does not depend on them 
\cite{Piguet:1995er}.

In addition, the action $\S^{\prime}$ satisfies the constraint
\eq
\int d^{4}x\; \left( \fud{\S^{\prime}}{c} + 
[\bar{c},\fud{\S^{\prime}}{b}] + \la \fud{\S^{\prime}}{L}\right) 
=\D^{cl}\ ,
\eqn{gheq}
where
\eq
\D^{cl} = \int d^{4}x\; \left ( [c,L ] +[D,\phi] + 
[\Omega^{\m}.A_{\m}] + [\xi^{\m},\psi_{\m}] + 
[\rho,\overline\phi] + [\tau,\eta] +  \half [ 
B^{\m\n},\chi_{\m\n}] \right )
\eqn{gheqbreak}
is a classical breaking. Eq. \equ{gheq} is known as the ``ghost 
equation'', and is a common feature of all gauge field theories in 
the Landau gauge \cite{Blasi:1991xz}.

Since the ordinary ST identity \equ{quantumst} is not affected by 
anomalies, and $\left(\s(x,)\la(x)\right)$ 
enter as a BRS doublet, the classical 
relation \equ{stidprime} can be extended at the quantum level, and we 
can write
\eq
\SS^{\prime}(\G^{\prime}) = 0\ ,
\eqn{quantumstidprime}
where $\G^{\prime}$ is the 1PI generating functional 
$\G^{\prime}=\S^{\prime} + O(\hbar)$. 

Similarly, it is easy to prove that the ghost equation \equ{gheq} is 
free of anomalies, and it holds for $\G^{\prime}$, too.

Differentiating the ST identity \equ{quantumstidprime} with respect 
to the external source $\la(x)$ and setting 
$\s(x)=\la(x)=0$\ , we obtain
\eq
\left.\fud{}{\la(x)}\SS^{\prime}(\G^{\prime})\;  \right |_{\s=\la=0} 
=0\ ,
\eqn{16}
that is
\eq
\left. b_{\G^{\prime}} \fud{\G^{\prime}}{\la(x)}\; 
\right |_{\s=\la=0} = 
\left.\fud{\G^{\prime}}{\s(x)} \;\right |_{\s=\la=0} \ ,
\eqn{17}
which means that the relation \equ{rel} between the composite 
operators $\Omega_{\phi^{2}}$ \equ{omfiquad} and $\Omega_{c^{3}}$ 
\equ{omcicub} holds 
true also at the quantum level~:
\eq
\Tr \phi^{2}(x)\cdot\G = b_{\G} \left[ \Tr \left( 
-\frac{c\phi}{\om^{2}} + \frac{c^{3}}{3\om^{4}}\right 
)(x)\cdot\G\right]\ .
\eqn{q}

As a consequence of the ghost equation \equ{gheq}, the quantum 
insertion \\\mbox{$ \Tr \left( 
-\frac{c\phi}{\om^{2}} + \frac{c^{3}}{3\om^{4}}\right 
)(x)\cdot\G$} has vanishing anomalous dimension. Indeed, in the 
Landau 
gauge the composite operators given by polynomials of the 
Faddeev--Popov ghost field $c(x)$ do not renormalize, whereas it is 
the differentiated ghost $\partial_{\m} c(x)$ which has quantum 
relevance \cite{Piguet:1995er,Blasi:1991xz}~:
\eq
\CC \left[ \Tr \left( 
-\frac{c\phi}{\om^{2}} + \frac{c^{3}}{3\om^{4}}\right 
)(x)\cdot\G\right] =0\ ,
\eqn{noandimccub}
where $\CC$ is the Callan--Symanzik (CS) operator
\eq
\CC \equiv \m\pad{}{\m} + \hbar \b_{g}\pad{}{g} - \hbar 
\sum_{\varphi}\gamma_{\varphi}\NN_{\varphi}\ ,
\eqn{cs}
and
\eq
\CC \G =0\ .
\eqn{cgamma}
In \equ{cs} $\b_{g}$ is the $\b$--function of the unique coupling 
constant $g$, $\g_{\varphi}$ is the anomalous dimension of the 
generic 
field $\varphi$, the operator $\NN_{\varphi}$ is the counting operator
\eq
\NN_{\varphi} = \int d^{4}x\; \varphi \fud{}{\varphi}\ ,
\eqn{18}
and $\m$ is the renormalization scale which appears in the 
$g$--normalization condition, for instance that which fixes the 
transverse 
part of the two--point function
\eq
\left. \frac{d}{dp^{2}}\G^{T}(p^{2})\;\right |_{p^{2}=-\m^{2}} = 
-\frac{1}{g^{2}}\ .
\eqn{19}
The Callan--Symanzik operator $\CC$ \equ{cs} commutes with the 
linearized ST operator~$b_{\G}$
\eq
[ \CC, b_{\G}] =0\ ,
\eqn{20}
hence, by applying $b_{\G}$ to \equ{noandimccub}, we have
\eq
\CC \left [ \Tr \phi^{2}(x)\cdot\G\right] =0\ ,
\eqn{noandimfiquad}
which completes the proof of vanishing anomalous dimension for the 
quantum insertion $\Tr \phi^{2}(x)\cdot\G$.

It is also immediate to verify that the integrated quantum insertions 
\eq
\int d^{4}x\; \WW^{4}\; \Tr \phi^{2}\cdot\G
\eqn{w4fiquad}
and
\eq
\int d^{4}x\; \WW^{4} \;  \Tr \left( 
-\frac{c\phi}{\om^{2}} + \frac{c^{3}}{3\om^{4}}\right )\cdot\G
\eqn{w4ccub}
have vanishing dimensions too, \ie
\begin{eqnarray}
    &&\CC \int d^{4}x\; \WW^{4}\; \Tr \phi^{2}\cdot\G =0 
    \label{noandimw4fiquad}\\
   && \CC \int d^{4}x\; \WW^{4} \;  \Tr \left( 
        -\frac{c\phi}{\om^{2}} + \frac{c^{3}}{3\om^{4}}\right 
	)\cdot\G =0\ .\label{noandimw4ccub}
\end{eqnarray}	
In fact, the same procedure used to show \equ{noandimfiquad} can be 
repeated, by coupling two global parameters to the integrated 
composite operators \equ{w4fiquad} and \equ{w4ccub}. By using the 
ghost equation \equ{gheq}, which holds unaltered, and the fact that, 
on 
integrated functionals, the operators $b_{\G}$ and $\WW_{\m}$ 
commute~\equ{b-W-dec}, one can play with 
the CS operator and reach the desired 
results \equ{noandimw4fiquad} and \equ{noandimw4ccub}.
%*******************************************************

%*****************************************************
\section{The Non-Renormalization Theorem for the beta 
Function}\label{sez6}

The starting point is the classical relation, deriving from the 
expression  of the twisted $N=2$ SYM action \equ{w4action},
at $\varepsilon^\m \equiv 0$
\eq
\pad{\S}{g} = \frac{2}{3g^{3}} \int d^{4}x  \;
\WW^{4} \;\Tr \frac{\phi^{2}}{2}
+ b_\S \int d^4x\; \Tr ( \phi D - \xi^\m \psi_\m )
\ .
\eqn{21}
The above expression can be written
\eq
\pad{\S}{g} = \frac{1}{g^{3}} b_{\S} \Xi\ ,
\eqn{dsdg}
with $\Xi$ given by
\eq
\Xi = \frac{2}{3} \int d^{4}x\;  \WW^{4}\; \Tr \left ( 
-\frac{c\phi}{\om^{2}} + \frac{c^{3}}{\om^{4}}\right) 
+ g^3 \int d^4x\; \Tr (\phi D - \xi^\m\psi_\m)
\ .
\eqn{22}
Our first intermediate task is the quantum extension of \equ{dsdg}. 
In 
order to do this, consider the action, modified by the introduction 
of 
a parameter $\a$
\eq
\S^{\prime} = \S + \frac{\a}{g^{3}}\Xi\ ,
\eqn{mods}
which is easily seen to satisfy the identity
\eq
\SS(\S^{\prime})+\a\pad{\S^{\prime}}{g}=0\ .
\eqn{modst}
The fact that $\a$ and $g$ transform as a BRS doublet, allows us to 
implement \equ{modst} at the quantum level
\eq
\SS(\G^{\prime})+\a\pad{\G^{\prime}}{g}=0\ .
\eqn{modqst}
Deriving \equ{modqst} with respect to $\a$ and setting $\a=0$\ ,
we find
\eq
\left. b_{\G^{\prime}}\pad{\G^{\prime}}{\a}\right|_{\a=0} = 
\left.\pad{\G^{\prime}}{g}\right|_{\a=0}\ .
\eqn{dgdg}
Now, at the classical level, from \equ{mods} we have
\eq
\pad{\S^{\prime}}{\a} = \frac{1}{g^{3}} \Xi\ ,
\eqn{23}
and $\Xi$ consists of two terms: $\int d^{4}x\;  \WW^{4} \Tr \left ( 
-\frac{c\phi}{\om^{2}} + \frac{c^{3}}{\om^{4}}\right)$, which has 
vanishing anomalous dimension, due to the nonrenormalization theorem 
\equ{noandimw4ccub};
and $g^3 \int d^4x\;\Tr (\phi D -\xi^\m\psi_\m)$,
which is purely classic, being a functional linear 
in the quantum fields $\phi(x)$ and $\psi_\m(x)$.

Consequently, we can extend \equ{23} to the quantum level
\eq
\pad{\G^{\prime}}{\a} = \frac{1}{g^{3}} \Xi\cdot\G^{\prime}
\eqn{24}
and, from the relation \equ{dgdg}, we have
\eq
\pad{\G}{g} = \frac{1}{g^{3}} b_{\G} \left ( \Xi\cdot\G\right)
\ ,
\eqn{25}
or
\eq
\pad{\G}{g} = \frac{2}{3g^{3}} \int d^{4}x\; 
\left(\WW^{4}\Tr\phi^{2}\right)\cdot\G 
+b_\G\int d^4x\;\Tr (\phi D -\xi^\m\psi_\m)\cdot\G
\ ,
\eqn{interaim}
which realizes our intermediate aim.

All the tools to prove the 1-loop finiteness of the beta function of 
the coupling constant are now at our disposal.

The 1PI generating functional of the theory $\G$ satisfies the CS 
equation \equ{cgamma}, or explicitly
\eq
\m\pad{\G}{\m} + \hbar \b_{g} \pad{\G}{g} - \hbar 
\sum_{\varphi}\g_{\varphi}\; b_{\G} (\Omega\cdot\G) = 0\ ,
\eqn{csgamma}
where we used the fact that the anomalous dimensions belong to the 
BRS--trivial sector of the counterterms. Deriving \equ{csgamma} with 
respect to the coupling constant $g$ we have
\eq
\m\pad{}{\m} \pad{\G}{g} + \hbar \pad{\b_{g}}{g}\pad{\G}{g} + \hbar 
\b_{g} \pad{}{g} \pad{\G}{g} - \hbar 
b_{\G}(\Omega^{\prime}\cdot\G)=0\ ,
\eqn{cs1}
with
\eq
\Omega^{\prime}\cdot\G \equiv 
\sum_{\varphi} 
\left ( 
\pad{\g_{\varphi}}{g}\Omega\cdot\G + \g_{\varphi} 
\pad{}{g}\Omega\cdot\G \right 
)\ .
\eqn{omega''}
It is important to emphasize that the quantum insertion 
$\Omega\cdot\G$, and 
hence $\Omega^{\prime}\cdot\G$, coming from the expression of 
counterterms, 
are analytic in the $\om$--parameter, as already remarked in 
section 4.

Substituting in \equ{cs1} the expression obtained in \equ{interaim} 
for the quantum insertion $\pad{\G}{g}$, and collecting all the exact 
terms at the r.h.s., we find
\eq
\CC \left[ \frac{1}{3g^{3}} \int d^{4}x\; \left( \WW^{4} \Tr 
\Phi^{2}\right) \cdot \G \right] + \hbar \pad{\b_{g}}{g} \left[ 
\frac{1}{3g^{3}} \int d^{4}x\; \left( \WW^{4} \Tr 
\Phi^{2}\right)\cdot\G\right] = \hbar 
b_{\G}(\Omega^{\prime\prime}\cdot\G)\ ,
\eqn{cs2}
where $\Omega^{\prime\prime}\cdot\G$ is 
again analytic in the parameter $\om$, being 
directly related to the expression of the classical action and of the 
counterterms. Remembering now the result \equ{noandimw4fiquad}, 
stating that the quantum insertion 
$\int d^{4}x\; \left(\WW^{4}\Tr\phi^{2}\right)\cdot\G$ has vanishing 
anomalous dimension, Eq. \equ{cs2} becomes
\eq
\left ( - \frac{3}{g^{4}}\b_{g} + \frac{1}{g^{3}} 
\pad{\b_{g}}{g}\right) \frac{1}{3} \int d^{4}x\; 
\left(\WW^{4}\Tr\phi^{2}\right)\cdot\G = 
b_{\G}(\Omega^{\prime\prime}\cdot\G)\ .
\eqn{cs3}
Both sides of \equ{cs3} can be written as cocycles, since it is 
evident that
\eq
\int d^{4}x\; 
\left(\WW^{4}\Tr\phi^{2}\right)\cdot\G = b_{\G} \int d^{4}x\; 
\left[ \WW^{4} \Tr \left( -\frac{c\phi}{\om^{2}} + 
\frac{c^{3}}{3\om^{4}}\right)\right]\cdot\G\ ,
\eqn{26}
which derives immediately from \equ{q}.

We thus have
\eq
\left ( - \frac{3}{g^{4}}\b_{g} + \frac{1}{g^{3}} 
\pad{\b_{g}}{g}\right) b_{\G} \int d^{4}x\; 
\left[ \WW^{4} \Tr \left( -\frac{c\phi}{\om^{2}} + 
\frac{c^{3}}{3\om^{4}}\right)\right]\cdot\G = b_{\G} \sum_{k} 
\om^{k}\; (\Omega^{\prime\prime}_{(k)}\cdot\G)\ .
\eqn{27}
That is, we have an equality between an analytic function of $\om$, 
in the r.h.s., and a non--analytic function on the l.h.s. The only 
solution is that the function $\b_{g}$ satisfies the differential 
equation
\eq
\frac{3}{g} \b_{g} + \partial_{g}\b_{g} = 0\ ,
\eqn{28}
\ie
\eq
\b_{g} = K\; g^{3} \ \ \ \mbox{(K constant).}
\eqn{result}
Eq. \equ{result} is our result. It states that, to all orders of 
perturbation theory, the $\b$--function of the coupling constant 
receives contributions only to the 1--loop order.

The exact statement is that a renormalization scheme does exist such 
that Eq.~\equ{result} is true, namely that all higher corrections 
vanish. Such scheme can be identified with the requirement that the 
insertion $\pad{\G}{g}$ can be defined exactly as expressed in Eq. 
\equ{interaim}, which is of the type
\eq
\pad{\G}{g} = \frac{1}{g^{3}} \int d^{4}x\; \Omega\cdot\G 
+b_\G(\D\cdot\G)\ .
\eqn{29}
% \eq
% \pad{\G}{g} = \frac{2}{3g^{3}} \int d^{4}x\; 
% \left(\WW^{4}\Tr\phi^{2}\right)\cdot\G 
% +b_\G(\D\cdot\G)\ .
% \eqn{29}

%*******************************************************

%************************************************************
\section{Summary}\label{sez7}
%************************************************************
In this paper we gave an algebraic, regularization independent, proof 
of the nonrenormalization theorem concerning the vanishing above one 
loop of the perturbative 
$\b$-function of $N=2$ SYM theory. The key points of our approach
have been the followings~:
\begin{description}
    \item[The twist~:] by means of the twisting procedure, we passed 
    from $N=2$ Super Yang--Mills theory to the equivalent 
    Topological Yang--Mills theory formulated on ${\real}^4$.
    \item[$\Tr\phi^{2}$~:] we were able to write the classical action 
    of the twisted theory as~\equ{w4action}
    $$\S \approx -\frac{1}{3g^{2}} \WW^{4} \int d^4x\;\Tr 
    \frac{\phi^{2}}{2}\ ,$$
    modulo a trivial BRS cocycle;
    \ie we related the classical action to $\Tr \phi^{2}$, which is a 
    gauge invariant  polynomial in the scalar field $\phi(x)$ of the 
    (untwisted) $N=2$ gauge supermultiplet.
    \item[Nonrenormalization of $\Tr\phi^{2}$~:] we have proved that 
    $\Tr\phi^{2}$ has vanishing anomalous dimension.
    \item[Callan-Symanzik equation~:]a wise manipulation of the 
    Callan-Symanzik equation implied that the beta function of the 
    unique coupling constant of the theory satisfies, to all orders of 
    perturbation theory, the following differential equation~\equ{28}
    $$\frac{3}{g} \b_{g} + \partial_{g}\b_{g} = 0\ ,$$
    which yields the result~\equ{result}
    $$\b_{g} = K\; g^{3}\ ,$$
    \ie the $\b$ function has only one loop contribution, in the 
    scheme defined by~\equ{29}.
\end{description}
%**************************************************************
\vspace{12mm}

\noindent   {\bf Acknowledgments}: It is a pleasure to thank Carlo 
Maria Becchi, Beppe Bandelloni and Nico Magnoli 
for discussions and for their 
comments. A.B.,  N.M. and A.T. would like 
to thank UERJ, where most of this 
work has been done, and the people of CBPF (Rio de Janeiro) 
for the warm 
hospitality and the friendly athmosphere. 
The financial support of the FAPERJ, Funda{\c c}{\~a}o de 
Amparo a Pesquisa do Estado do Rio de Janeiro, and the SR2--UERJ are 
gratefully acknowledged. 

%*****************************************************************

%****************************************************************
%\appendix
%\renewcommand{\theequation}{\Alph{section}.\arabic{equation}}
%\renewcommand{\thesection}{\Alph{section}}
%\setcounter{equation}{0}
%\setcounter{section}{1}
%\section*{Appendix}

\end{document}